\documentstyle[12pt,epsf]{ioplppt}
\begin{document}
\submitted
\jl{6}   

\def\trK{\hbox{Tr}K}
\date{\today}
\title{A fully $(3+1)$-D {R}egge calculus model of the {K}asner
cosmology}[Regge model of the {K}asner cosmology]

\author{Adrian P. Gentle\footnote{Permanent address: 
                                Department of Mathematics, 
                                Monash University, 
                                Clayton, Victoria 3168, Australia.
                                Email: adrian@newton.maths.monash.edu.au}
        and Warner A. Miller\footnote{wam@lanl.gov}}
\address{Theoretical Division (T-6, MS B288),\\ Los Alamos National 
         Laboratory, Los Alamos, NM 87545, USA.}

\begin{abstract}
We describe the first discrete-time 4-dimensional numerical
application of Regge calculus. The spacetime is represented as a
complex of 4-dimensional simplices, and the geometry interior to each 
4-simplex is flat Minkowski spacetime.  This simplicial spacetime is
constructed so as to be foliated with a one parameter family of
spacelike hypersurfaces built of tetrahedra.  We implement a novel
two-surface initial-data prescription for Regge calculus, and provide
the first fully 
4-dimensional application of an implicit decoupled evolution scheme
(the ``Sorkin evolution scheme'').  We benchmark this code on the
Kasner cosmology --- a cosmology which embodies generic features of the
collapse of many cosmological models.  We (1) reproduce the continuum
solution with a fractional error in the 3-volume of  
$10^{-5}$ after $10000$ evolution steps, (2) demonstrate stable
evolution, (3) preserve the standard deviation of spatial homogeneity
to less than $10^{-10}$ and (4) explicitly display the existence of
diffeomorphism freedom in Regge calculus.   We also present the
second-order convergence properties of the solution to the continuum.
\end{abstract}

\setcounter{footnote}{0}


\section{Regge calculus as an independent tool in general relativity}
\label{sec:intro}

In this paper we describe the first fully $(3+1)$-dimensional
application of Regge calculus \cite{regge61,williams92} to general
relativity. We develop an initial-value prescription based on the
standard York formalism, and implement a 4-stage parallel evolution
algorithm.  We benchmark these on the Kasner cosmological model. 

We present three findings.  First, that the Regge solution exhibits
second-order convergence of the physical variables to the continuum
Kasner solution.  Secondly, Regge calculus appears to have a complete
diffeomorphic structure, in that we are free to specify three shift and
one lapse condition per vertex.  Furthermore, the four corresponding
constraint equations are  conserved, to within a
controllable tolerance, throughout the evolution. Finally, the
recently-developed decoupled parallel evolution scheme \cite{committee}
(the ``Sorkin evolution scheme'') yields stable evolution.

Although we have taken just the first few steps in developing a
numerical  Regge calculus programme, every indication (both numerical
and analytic) suggests that it will be a valuable tool in the study of
gravity.   Our numerical studies, together with analytic
results \cite{miller86c}  should put to rest 
some of the recent concerns about Regge calculus
\cite{brewin95,miller95,reisenberger96} --- 
it does  appear to be a viable approximation to general relativity.

Einstein described gravitation through the curvature of a
pseudo-Riemannian manifold.  Regge calculus, on the other
hand, describes gravity through the curvature of a piecewise flat
simplicial pseudo-Riemannian manifold.  The
fundamental platform for Regge calculus is a lattice spacetime,
wherein each lattice cell is a simplex endowed with a flat Minkowski
geometry \cite{williams92,wheeler64,mtw}.  The physical and geometric
basis of 
Regge calculus distinguishes it from all other discretizations of
general relativity.   One applies the principles of Einstein's theory
directly to the simplicial geometry in order to form the curvature,
action and field equations \cite{miller97}.    This is in stark
contrast to the finite 
difference approach, where one starts with a representation of the
continuum field equations and proceeds to discretize them
over a grid of points.  The goal of our work is to evaluate
the relative strengths and weaknesses of Regge calculus.

The geometrically transparent nature of Regge calculus should be
useful in the interpretation of simulations.  One example of this is
the Kirchoff-like form of the contracted Bianchi identity
\cite{miller86c,kheyfets90}  in Regge calculus.   All edges in
the lattice geometry, in a strict sense,  carry a flow of
energy-momentum, and the sum of these flows at each vertex is
zero.   This single example illustrates that every term in every Regge
equation has a clear geometric interpretation.

We choose to benchmark our code on the Kasner cosmology because it has
a well defined solution with symmetries.  More importantly, however, 
the Kasner model is a prototype of the Belinsky-Khalatnikov-Lifschitz
mixmaster oscillation generic to all crunch cosmologies
\cite{belinsky70}. 
Unlike all previous Regge simulations, we impose no symmetries
on the model --- it is free to exhibit all dynamical degrees of
freedom. The symmetries of the Kasner cosmology are encoded in the
freely specifiable portion of the initial data, and the full Regge
initial-value problem is then solved.  The solution obtained displays
encouraging agreement with the continuum Kasner solution, and this 
agreement is  found to be remarkably well preserved during
evolution.  The code used in these calculations is applicable to
$S^3\times R^1$ spacetimes, as well as the $T^3\times R^1$  topology
presented here. 

There are two major areas of active research in numerical relativity
today. The first effort deals with gravitational wave generation by
astrophysical processes.  The Laser Interferometric Gravity Wave
Observatory, which should be operational within a few years, will
require a set of gravity wave templates from colliding and coalescing
black holes. The second area of research deals with the structure of
cosmological singularities in inhomogeneous cosmologies.  The gravity
wave problem is dominated in many ways by boundary conditions, whilst
the cosmological singularity problems have no boundary.
We have chosen to focus our initial application of Regge
calculus toward issues in inhomogeneous cosmology.

In this paper we will develop the simplicial lattice used in the
calculations (section \ref{sec:kinematics}), and then briefly describe
the geometric and dynamical structure of Regge calculus (section
\ref{sec:dynamics}).  These ideas will be applied to motivate 
a York-style two-surface initial-value prescription for Regge
calculus (section \ref{sec:ivp}), which is then specialized to the
Kasner cosmology (section \ref{sec:kasner}).
With initial data in hand, we proceed to the time evolution
problem in section \ref{sec:evolution}, and apply a four-step
implementation of the Sorkin evolution 
scheme.   We examine the convergence of the Regge solution to
the continuum, and confirm the existence of simplicial gauge freedoms 
by demonstrating that the constraint equations are satisfactorily
preserved during evolution.  Finally, in section
\ref{sec:conclusion}, we discuss our future plans for numerical Regge
calculus. 

\section[]{Kinematics and the Quantity Production Lattice}
\label{sec:kinematics}

The first issue one must address when beginning a Regge calculus
simulation is the choice of lattice structure.   How do we choose to
represent a spacetime geometry with a lattice geometry?  We are aided
in our decision by the following four guiding principles
\begin{itemize}
\item[\bf{(i)}] We want the simplicial spacetime foliated into 3-dimensional
tetrahedral spacelike hypersurfaces.  Although these hypersurfaces
will be geometrically distinct, their lattice structure should be
identical.   The spacetime sandwiched between any two of these
surfaces should be decomposed into simplexes.
\item[\bf{(ii)}] Each spacelike hypersurface should have the topology
of a 3-torus ($T^3$).
\item[\bf{(iii)}] The simplicial structure sandwiched between
two adjacent hypersurfaces must be consistent with the recently
proposed Sorkin evolution scheme \cite{committee}. 
\item[\bf{(iv)}] The lattice connectivity, or local topology, of the
3-geometries should be maximally homogeneous. In other words, the
lattice structure at one vertex should be identical to all others.
\end{itemize}
These four guiding principles have led us to the choice of lattice
described below.  It is in this sense that we have introduced a $(3+1)$
split of spacetime in Regge calculus.

The fundamental geometric element in Regge calculus is the
four-simplex, consisting of five vertices, ten edges, ten
triangular hinges, and five tetrahedra.  In this section we build
the four dimensional simplicial lattice by constructing a 
simplicial three geometry, and then carry the triangulation of
$T^3$ forward in time, to obtain a simplicial manifold with topology
$T^3\times R^1$.

The best tetrahedral subdivision of $T^3$ that we know is based on
isosceles tetrahedra, and will be referred to as the Quantity
Production Lattice (QPL) \cite{miller86b}.  A major advantage of this
lattice is that it is easily refined whilst maintaining local
homogeneity, in the sense that the connectivity at each vertex is the
same.  The QPL consists of isosceles tetrahedra only in flat Euclidean
space, with slight variations in the edge lengths introducing
curvature about edges in the three-lattice.  

\begin{figure}
\centerline{\epsffile{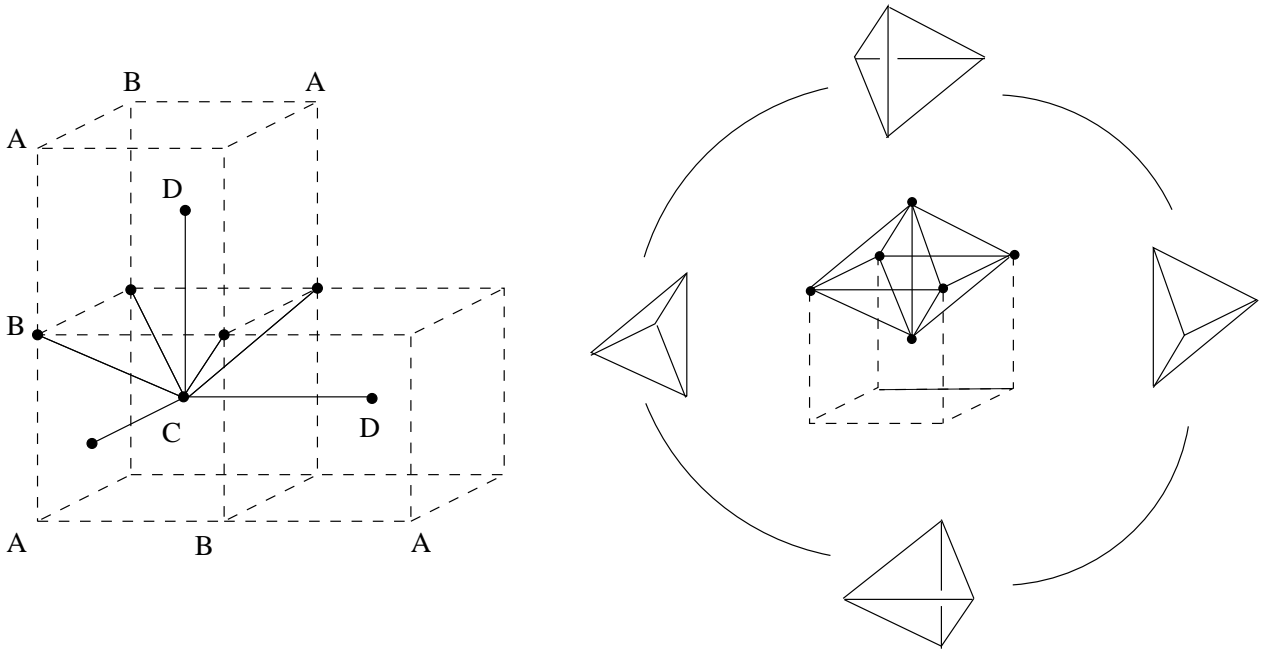}}
\caption{\protect\small{\em{Two different views of the
Quantity Production Lattice.  a) The vertex data structure, which when
carried from vertex to vertex generates the three-geometry.  Several
A-, B-, C-, and D-type vertices are also shown, and these are defined
in the text below.   b)  The local
lattice structure about a single leg.  The four tetrahedra
piercing a single face of the original cubic lattice are shown
``exploded off''. }}}
\label{fig:qpl}
\end{figure}

The $T^3$ QPL may be constructed from a single
cube. Begin by identifying opposing faces, thus fixing the global
topology. 
Subdivide the cube into smaller cubes, until the desired resolution is
obtained, and then introduce a new vertex at the centre of
each small cube.  Joining these centred vertices together yields a
new cubic lattice which pierces the faces of the original.   Finally,
join the centred vertices to each of the eight vertices of
the original cube in which they reside, creating four tetrahedra
through each face of the original cubic lattice.   This completes the
construction of the QPL. Figure \ref{fig:qpl} displays some of the
local structure of the resulting lattice. 

Fourteen edges emanate from each and every vertex in the QPL. Six 
legs lie along cube edges, and
8 diagonal edges join the two cubic lattices together.
These two types of edges will be refered to as cube-aligned
and diagonal edges, respectively.  The QPL, as outlined here, is
isomorphic to the right-tetrahedral lattice, which can be formed by
adding face and body diagonals to a cubic grid.  Despite this, we have
found that  assigning  edge lengths using the right tetrahedral
approach leads to minor numerical difficulties, due to degeneracy in
the first derivatives of the Regge equations.  This has perhaps impeded
previous work with $(3+1)$ formulations of Regge calculus
\cite{miller96}. 

We now have in hand the simplicial three geometry, and turn to the
construction of the four geometry in which this initial slice is to be
embedded.  The four geometry is constructed by dragging 
vertices forward, one by one, until the whole initial surface
has been replicated, and the intervening region filled with
four-simplices.  We refer to the resulting lattice as a 
Sorkin triangulation \cite{sorkin75,tuckey92}. 

Another useful approach to the construction of the four geometry is
the ``vertex data structure'', whereby each vertex in the
lattice has, in so far as is possible, an identical local structure.
In general, the Sorkin approach to the evolution problem is
inconsistent with such a lattice.   The major advantage of Sorkin
triangulations over a four geometry built using a vertex data
structure  is that it allows the use of an order $N_v$ (number
of vertices in each three-geometry), parallel, four-step evolution
algorithm.  That is, one quarter of the vertices in the three-geometry
may be evolved together, in parallel.  

We construct the Sorkin triangulation as follows.  Identify four
classes of vertices in the QPL, 
two types on the original cubic structure, and the remaining two on
the cube-centred lattice .  The vertex types are defined as 
\begin{center}
\begin{itemize}
\item[\sl{A-type}:] Half the vertices on the original cubic lattice,
        sharing no common edges.
\item[\sl{B-type}:] The remaining vertices on the original cubic lattice.  
\item[\sl{C-type}:] Half the vertices on the cube--centred lattice,
not sharing common edges.
\item[\sl{D-type}:] Remaining vertices on the cube--centred lattice.
\end{itemize}
\end{center}
These identifications are indicated in figure \ref{fig:qpl}. 
We now  construct the four geometry from the simplicial
three geometry.
 
\begin{figure}
\centerline{\epsfxsize=3truein\epsfbox{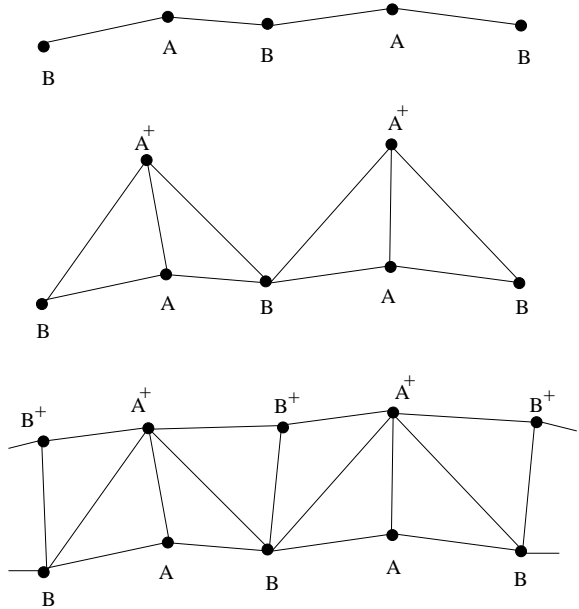}}
\caption{\protect\small{\em{The $(1+1)$-dimensional analogue of the
Sorkin scheme, applied to a region of the lattice.
    a) On the initial hypersurface, we identify A- and
B-type vertices, the only requirement being that no two vertices of a
given type  share an edge. In $(3+1)$ dimensions, the lattice
breaks into four classes of vertices under this requirement.  b) All
A-type vertices are carried forward together, creating (n+1)-simplices
``above'' each n-simplex in the initial slice.   c) The B-type
vertices are dragged forward.  This completes the $(1+1)$-D evolution
procedure, replicating the original hypersurface and triangulating the
intervening region.}}}
\label{fig:1+1}
\end{figure}

Begin by selecting a single A-type vertex, $A_k$ say, and drag it
forward in time, to create a new vertex $A_k^+$. In the process we
introduce 14 brace edges 
over the 14 legs emanating from $A_k$ in the $l^{th}$ spacelike
hypersurface ($\Sigma_l$),
together with a vertical (timelike) leg joining $A_k$ to $A_k^+$.   
This timelike edge is taken small enough to ensure that the 
brace edges are spacelike.
Each brace edge joins the newly evolved vertex $A_k^+$ to the
corresponding vertices in the previous slice 
which share an edge with $A_k$.
Figure \ref{fig:1+1} shows the situation in $(1+1)$ dimensions.  From
the figure it is clear that ``evolving'' an A-type vertex creates a
2-simplex  (triangle) over each 1-simplex (edge) meeting at $A_k$.  In
$(3+1)$ dimensions, the procedure creates a 4-simplex above each
tetrahedron (3-simplex) which shares the vertex $A_k$.  Having
identified the 4-simplices in this way, we may tabulate the new
tetrahedra and triangles which they contain.  This process can be
repeated on all A-type vertices simultaneously, since none of these
vertices share a common edge.  

Now consider a single B-type vertex.
Dragging it forward yields 8 brace legs between
slices, one over each of the 8 diagonal legs in $\Sigma_l$.   The
original vertex is also joined to 6 A-type vertices which have already
been dragged forward, so joining the new vertex to these creates 6
spatial legs on $\Sigma_{l+1}$.  A vertical edge
joining the new vertex to its counterpart on the original slice is
also created.  Again, all B-type
vertices can be dragged forward in parallel, and this process creates
a set of 4-simplices in a manner similar to the A-type evolution step.

Next the cube--centred C-type vertices.   Each of these creates 6
brace legs 
between slices, since each C-type vertex in $\Sigma_l$ is connected to
6 D-type vertices, which have yet to be carried forward.  Also created
are 8 spatial diagonal legs on $\Sigma_{l+1}$, joining the C-type to
surrounding A- and B-type vertices, together with a single timelike
edge.   Finally, the D-type vertices are dragged forward, creating
14 new spatial legs on $\Sigma_{l+1}$, together with a single timelike
leg.   

In this way the initial surface is carried forward in time, 
constructing an identical three geometry on $\Sigma_{l+1}$,
and filling the intervening region with four-simplices.  The resulting
$(3+1)$ Regge spacetime is consistent with our four guiding
principles. 

\section{Dynamical structure of Regge calculus}
\label{sec:dynamics}

In section \ref{sec:kinematics} we developed the kinematic
structure of the simplicial spacetime.  We now apply the Regge
form of Einstein's field equations to the lattice.

Regge \cite{regge61,miller97} constructed the simplicial form of the
Hilbert 
action, and obtained what have become known as the Regge equations via
a variational principle.  The independent variables in this approach
are the lattice legs. Each edge in the lattice has
associated with it a simplicial equivalent of the Einstein equations,
which take the form \cite{regge61,miller86c}
\begin{equation}  \label{eqn:regge}
  \frac{2L}{V^*_L}  \sum _h \epsilon _h \frac{\partial A_h}
        {\partial  l^2_j} = 0, 
\end{equation}
where the summation is over all triangles $h$ which hinge on the leg
$l_j$.    The defect $\epsilon_h$ about $h$ is the angle
that a vector rotates when parallel transported around the hinge, 
$A_h$ is the area of the hinge, and $V^*_L$ is the 3-volume of the
Voronoi cell dual to edge $L$. 
Equation \ref{eqn:regge} is the Regge calculus analogue of
$G_{\mu\nu}=0$. 

When applied to our lattice, there are three types of Regge equation,
depending on the 
type of edge in the lattice it is associated with.  The equations
associated with edges in a spacelike hypersurface may be viewed as
``evolution''-type equations, since they couple information on that
surface to the edges lying above and below.  This is the Regge
equivalent of the second order evolution equations in the ADM $(3+1)$
split \cite{mtw}.

Equations associated with brace edges, those lying between successive
spacelike hypersurfaces, only involve quantities on and between the 
two slices.   In this sense they are ``constraint'' or first order
equations. Similarly, the ``timelike Regge equations'' obtained by
varying the vertical edges between two spacelike hypersurfaces are
first order equations.

The Einstein tensor in the continuum is a rank 2 symmetric tensor, and
hence has two indices. E. Cartan
has provided  us with a geometric interpretation of this tensor and
its 
components. He showed that the Einstein tensor is expressible as a
double dual of the sum of moments of rotation \cite{mtw} over the
2-dimensional faces of a 3-volume.   Thus one index can be interpreted
as the orientation of that 3-volume, and the other can be associated
with the orientation of the dual of the sum of moments of rotation.

The Cartan approach
provides a derivation of the Einstein equations independent of the
action, and has been successfully  applied to Regge calculus
\cite{miller86c}.   The usual Regge equations are recovered as a sum
of moments of rotations.  Thus for each edge $L$ in the Regge spacetime
the voronoi 3-volume $V^*_L$ dual to the edge, and each of the moment of
rotation vectors over the 2-dimensional faces of $V^*_L$ are parallel and
directed along $L$, yielding a single Regge equation per edge. 
It is in this sense that the Regge equation is the double projection
of the Einstein tensor along edge L (ie: $G_{LL}$).  
Since the timelike edges in our Kasner simulation are orthogonal to
the homogeneous spacelike hypersurfaces, the Regge equations
associated with these edges (the ``timelike Regge equations'') are the
canonical Hamiltonian constraint equations.  Similarly, the other
first-order Regge equations, associated with the brace edges, are the
only Regge equations that carry components of the momentum constraints
($G_{0i}$).  Thus we use three of these brace equations per vertex as
momentum constraints, and the timelike Regge equation as the
Hamiltonian constraint.

In continuum general relativity, the ten Einstein equations are
functions of the ten components of the metric tensor.  However, of
these ten metric components, only six functions per spacetime point
may be considered truly independent, since we have the freedom to
choose a system of co-ordinates.  Correspondingly, 
there are four relations per point amongst the Einstein equations
themselves, the contracted Bianchi identities \cite{mtw}.

A similar structure is found in Regge geometrodynamics.   Each vertex
in the lattice may be associated with  a unique set of $N$ edges, and 
each of these edges has a corresponding Regge equation.    However,
the simplicial form of the contracted Bianchi identity holds at each
vertex \cite{kheyfets90}, a total of four constraints per vertex.  So
there can only be $N-4$ truly independent edges per vertex, with the
remaining four edges representing simplicial gauge freedom. 
This corresponds to the lapse and shift freedom in continuum $(3+1)$
relativity \cite{mtw}.  The four redundant Regge equations per vertex
introduced in this process may be compared with the four constraint
equations in continuum $(3+1)$ general relativity. 

The finite rotations involved in the simplicial contracted Bianchi
identity do not commute, so these 
``identities'' are only approximate \cite{miller86c}.  
In the infinitesimal limit, however, it is expected
that the Regge Bianchi identities become exact in correspondence to
the continuum.  This failure to precisely
conserve the constraints is no worse than the failure of a finite
difference scheme to exactly conserve the numerical constraints.
We demonstrate this conservation of energy-momentum in section
\ref{sec:evolution}. 

\section[]{A York-type initial-value prescription for Regge calculus} 
\label{sec:ivp}

Before evolution can proceed, we must construct initial data
consistent with the Regge constraint equations.
All previous initial data constructed in Regge calculus has
been for the special case of a moment of time symmetry, where the
problem reduces to the requirement that the scalar curvature of the 
initial three geometry is zero \cite{wheeler64}.   
Successful time-symmetric Regge calculations include initial data  
for single and multiple black holes \cite{wong71,collins72},
Friedmann-Robertson-Walker and Taub cosmologies
\cite{collins73,brewin87,williams85}, and Brill waves on both
flat \cite{dubal89a} and black hole \cite{gentle96} backgrounds.

The major simplification of calculating initial data at a moment of
time symmetry is that the problem reduces to the calculation of the
purely three-dimensional scalar curvature, and does not involve
the four dimensional lattice structure \cite{wheeler64}.   In 
spacetimes of 
astrophysical interest there will not ordinarily be a moment of time
symmetry to simplify the initial-value problem.  Indeed, this is the
case in the Kasner class of cosmologies considered here.   
The discrete time Regge lattice neccesitates the construction of fully
four-dimensional, two-slice initial data (a ``thin sandwich
approach'').   We apply a novel 
two-slice initial-value formalism, to be described in more detail
elsewhere \cite{initial}, to construct initial data for the Kasner
cosmology in Regge calculus.

The two-surface approach outlined below is constructed to mirror, in
as far as possible, the York \cite{mtw,wheeler88} technique for
constructing 
initial data in the continuum, rather than the Belasco-Ohanian
\cite{belasco69} two-surface formulation.  The advantage of the York
approach is that 
it clearly delineates the true degrees of freedom of the gravitational
field from the embedding quantities.  York begins by
conformally decomposing the three-metric, 
\[
^{(3)}g_{ij} = \psi^4 \, ^{(3)}\tilde g_{ij}, 
\]
introducing the conformal factor $\psi$, and base metric $\tilde
g_{\mu\nu}$.  We are then free to specify \cite{mtw} the
\begin{itemize}
\item base metric $^{(3)}\tilde g_{ij}$ (``Where'').
\item momenta $\tilde \pi ^{TT}_{ij}$ (``How Fast''), and
\item The trace of the extrinsic curvature, $\trK$ (``When'').
\end{itemize}
Once these have been fixed, the only remaining
variables are the conformal factor $\psi$, and the gravitomagnetic
three-vector potential $W^i$.  The four constraint equations are 
used to calculate these quantities.

The key to our simplicial two-surface formulation of York's procedure
is the simplicial representation of the dynamical degrees of freedom
and embedding variables.  The identification presented below is by no
means unique. In other words, we provide a representation of the
degrees of freedom and not a diffeomorphically invariant, conformally
invariant representation of the dynamic degrees of freedom. To this
end, it is convenient to  perform a conformal decomposition on
each hypersurface, yielding leg lengths in $\Sigma_k$ ($k=0,1$) of
the form
\[
    l_{ij} = \psi_{ij} ^2 \, \tilde l_{ij}
\]
where $\tilde l_{ij}$ is the base leg between vertices $i$ and
$j$ lying in one hypersurface, the simplicial equivalent of the base
metric $^{(3)}\tilde 
g_{ij}$.  The conformal factor is defined on the vertices of the
3-lattice $\Sigma$, and is applied to the edge between vertices $i$
and $j$ using a centred, second order, approximation
\[
\psi _{ij}  =  \frac{1}{2} \left( \psi_i + \psi_j \right).
\]

Within each hypersurface, 14 edges emanate from
each vertex; 6 cube-aligned, together with 8
diagonal legs.  It is convenient at this stage to view the lattice,
and the three-lattice in particular, from the viewpoint of a vertex
data structure.  The three geometry can be constructed by applying the
generator shown in figure \ref{fig:qpl} to each vertex, which 
consists of 3 cube-aligned legs, together with 4 diagonal edges.
The four geometry sandwiched between the two
hypersurfaces contains a single brace leg over each of these 7 edges
per vertex, and so a similar decomposition into edges per vertex can
be obtained for the brace legs.  However, the exact structure of
these braces depends on the choice made  in the construction of the
four lattice, and in particular, each type of vertex (A,B,C,D) will have
a different arrangement of braces about it.

Since we are dealing with a full four-dimensional region of the
lattice, the initial-value problem requires that we specify 
lapse and shift, otherwise we would be unable to construct the
thin--sandwich analogue of $\trK$ and $\tilde \pi^{TT}_{ij}$.  These
quantities both depend on derivatives of the 3-metric, the lapse
function and the shift vector.
Our freedom to freely choose lapse and shift is linked to the
simplicial 
contracted Bianchi identities, discussed above \cite{miller86c}.
Although Regge 
calculus deals directly with geometric, co-ordinate independent
quantities, we must still specify how vertices on the initial three
geometry are pushed forward in time.   

The York prescription outlined above allows us to freely specify the 
base three-geometry (the ``where'' \cite{wheeler88}), $\trK$
(``when''), and the momentum conjugate to the true dynamical degrees
of freedom of the gravitational field (``how fast'').   

A two-surface formulation
of these ideas requires that the momenta terms be split across both
surfaces.   In direct analogy to the assignment of the base
three-metric, we specify freely all base legs on the initial
hypersurface $\Sigma_0$.   To avoid specifying everything on either
surface (and thus obtaining a Belasco-Ohanian style initial data
prescription \cite{belasco69}), we must not specify the conformal
factor $\psi$ on 
$\Sigma_0$.  The conformal factor at each vertex on $\Sigma_1$ is
freely specified instead, in analogy to the fixing of $\trK$. The
final step 
is to identify the Regge analogue of $\tilde \pi_{ij}^{TT}$.   We choose
the 4 diagonal base  legs per vertex on $\Sigma_1$ to represent this 
freedom, since, together with the lapse and shift choice, they enable
us to specify part of the change in the base ``metric'' across the
sandwich. 

Our two-surface formulation on the lattice, in the spirit of the
standard York decomposition, may be summarized as follows.  Specify
\begin{center}
\begin{itemize}
\item All 7 base edges $\tilde l_{ij}$ per vertex on $\Sigma_0$
(``Where''), 
\item The 4 diagonal base edges per vertex on $\Sigma_1$ (``How
Fast''),  
\item The conformal factor $\psi_{ij}$ at each vertex on $\Sigma_1$
(``When''), 
and 
\item 4 lapse and shift conditions per vertex (``How Fast'' and
``When''), 
\end{itemize}
\end{center}
which leaves the conformal factor $\psi$ on $\Sigma_0$, the 3
spatial 
cube-aligned edges per vertex on $\Sigma_1$, and the 4 braces 
per vertex which lie over diagonal edges as the true representations
of the geometric degrees of freedom.   These 8 unknowns per vertex may
be calculated by solving the 8 Regge constraint equations per vertex
which are available during the solution of the initial-value problem.
The constraints are associated with the 7 braces per vertex, together
with the timelike edge joining the vertex on $\Sigma_0$ to its
counterpart on $\Sigma_1$.

It is
possible to introduce sophisticated definitions of lapse and shift by
mirroring the 
continuum structure \cite{committee}, however a more natural approach is
to select appropriate legs in the lattice, and define them to
represent the simplicial gauge freedom.  In this spirit, we
define lapse to be the proper time measured along a timelike leg
joining a vertex to its counterpart on the next hypersurface, and the
shift to be a combination of three legs that uniquely fix the
location of the new vertex above the previous hypersurface.

Detailed definitions of shift differ as to which type of vertex (A,B,C
or D) we are considering.
The evolution of an A-type vertex  creates 14 brace
edges stretching between $\Sigma_l$ and the nascent hypersurface
$\Sigma_{l+1}$, together with a timelike ``Regge lapse'' edge.  Six
of these edges will lie above cube-aligned edges 
in $\Sigma_l$, and it is these braces
that we select as our simplicial shift freedom.  In particular, at a
given A-type vertex, the shift edges correspond to the brace legs
above the 3 cube-aligned edges in the vertex generator shown
in figure \ref{fig:qpl}.  This definition of shift is also
applied to the C-type vertices.

In the case of B- and D-type vertices, assigning the lengths of these
braces will not determine the position of the evolved vertex.
This can be seen in figure \ref{fig:1+1}, where assigning all brace
lengths will only rigidify the ``A''-type vertices, leaving the
``B''-type to flap.  This would result in an ill-conditioned system of
equations, since the location of the B-type vertices is completely
undetermined.   This $(1+1)$-dimensional example suggests that we 
define the shift edges on B- and
D-type vertices by assigning the lengths of the 3 cube-aligned edges
in $\Sigma_{l+1}$.    The three edges chosen are always those
associated with the vertex-by-vertex generator shown in figure
\ref{fig:qpl}. It is clear from this discussion that care must be
taken in choosing the simplicial counterparts of lapse and shift.

\section{Thin-sandwich initial-value approach applied to the Kasner
cosmology} 
\label{sec:kasner}

We now apply the initial-value formalism outlined above to the
Kasner class of cosmologies.  The metric for
these $T^3\times R^1$ vacuum solutions of Einstein's equations takes the
form 
\begin{equation}   \label{eqn:metric}
  ds^2 = - dt^2 + t^{2p_1} dx^2 + t^{2p_2} dy^2 + t^{2p_3} dz^2 
\end{equation}
where the unknown constants $p_i$ satisfy
\begin{equation}
 p_1 + p_2 + p_3 = p_1^2 + p_2^2 + p_3^2 = 1 .
\end{equation}

We construct the two surface initial data such that the initial
surface, $\Sigma_0$, is at $t=1$, and $\Sigma_1$ is at $t=1+\Delta
t$.  The base leg lengths on $\Sigma_0$ are assigned the flat-space
values obtained by setting $t=1$ in (\ref{eqn:metric}). This yields
\begin{equation}  \label{eqn:first}
\tilde l_{x_k} = \Delta x_k
\end{equation}
for the three cube-aligned base edges per vertex on $\Sigma_0$, and 
\begin{equation}
 \tilde l^2_{ij} = \frac{1}{4}\left( \Delta x^2 + \Delta y^2 + \Delta z^2
\right).
\end{equation}
for the 4 diagonal base edges per vertex.

The conformal factor on $\Sigma_1$ is the simplicial equivalent of 
setting $\trK$, and we choose to take 
\begin{equation}
\psi = 1 \quad \hbox{on} \quad \Sigma_1.
\end{equation}
The kinematic degrees of freedom correspond to the 4 diagonal legs per
vertex on $\Sigma_1$, and these are set in the same manner as the base
diagonal legs on the initial slice, using (\ref{eqn:metric}) evaluated
at $t=1+\Delta t$.  This yields
\begin{equation}   \label{eqn:next_leg}
 \tilde l^2_{ij} = \frac{1}{4}\left( t^{2p_1}\Delta x^2 
        + t^{2p_2} \Delta y^2 + t^{2p_3}\Delta z^2  \right).
\end{equation}

The only edges which remain to be fixed are our choice of shift and
lapse.  The lapse edge is assigned the squared proper time between
slices, 
\begin{equation}  \label{eqn:last}
    \tau_{ii^+} ^2 = - \Delta t^2, 
\end{equation}
and the squared lengths of the shift edges are obtained by applying a
power series  expansion along the continuum geodesics.
Accurate to third order in the
lattice spacing, the expansion between vertices $i$ and $j$ is
\begin{equation}  \label{eqn:series}
        l^2_{ij} = g_{\mu\nu} \Delta x_{ij}^\mu \Delta x_{ij}^\nu
                   + g_{\alpha\beta} \Gamma ^\beta_{\mu\nu}
                     \Delta x_{ij}^\alpha \Delta x_{ij}^\mu 
                     \Delta x_{ij}^\nu + \ldots
\end{equation}
where the continuum metric $g_{\mu\nu}$ is used to obtain $\Gamma
^\beta_{\mu\nu}$.  The spatial edge length assignments above are
identical 
to the expressions obtained from (\ref{eqn:series}), to third order.
In principle we could ignore the cubic term in the expansion, and
take the squared shift edge lengths to be  
\begin{equation}
l^2_{ij} = g_{\mu\nu} \Delta x_{ij}^\mu \Delta x_{ij}^\nu,
\end{equation} 
however it was found that the initial-value problem converged faster
using the higher order expansion.  The simplest approximation was 
tested, and found not to change the character of the initial-value
solution appreciably.    These expansions are 
used only whilst constructing initial-value data, and only when we
wish to compare with a continuum metric.

The shift conditions on A- and C-type vertices are applied to brace
legs lying above cube-aligned edges, and the geodesic-length 
approximation (\ref{eqn:series}) for these yields
\begin{equation} \label{eqn:shift1}
    d^2_{x_k} = - \Delta t ^2 + \Delta x _k^2 + p_k \Delta x_k^2 \Delta t.
\end{equation}
For the  B- and D-type vertices,  shift is applied to
the three cube-aligned base spatial edges per vertex in $\Sigma_1$,
where the series expansion takes the form
\begin{equation} \label{eqn:shift2}
    \tilde l_{x_k}^2 = - \Delta t ^2 + \left( 1 + \Delta t \right)
              ^{2p_k} \Delta x_k^2  .
\end{equation}

In all calculations shown in this paper, we use the QPL obtained by
two barycentric subdivisions of the original cube.  The three geometry
consists of  128 vertices, 896 legs, 1536 triangles and 768 tetrahedra
per spatial hypersurface.  The region contained between two
consecutive  spacelike hypersurfaces contains 3072 four-simplices.
Whilst this is a modest grid resolution by modern standards, the major
aim of this work is the evaluation of Regge calculus as a competitive
technique in $(3+1)$ numerical relativity;  we feel that the grid
is sufficient for this task, and experiments with increased spatial 
resolution do not effect the conclusions reached below.

The Kasner exponents $p_i$ enter into the
freely specifiable portion of the initial data through equations
(\ref{eqn:next_leg}), (\ref{eqn:shift1}) and (\ref{eqn:shift2}).  They
are not used at any other stage of the calculations, and in
particular, do not appear in the evolution problem.   The information
borrowed from the continuum solution in this way could, in principle,
be constructed within the Regge framework by seeking a solution with
the required symmetries.  Analytic calculations in which the Regge
lattice is forced to maintain the symmetries of the Kasner solution
have been undertaken \cite{lewis82,gentle95}.  These confirm that
the Regge equations produce the expected Einstein equations for the
Kasner cosmology to leading order in the limit of very fine
discretizations.   Any calculation must begin with a suitable ansatz
to determine the desired class of solutions; we have chosen the freely
specifiable portion of the initial data to closely model the Kasner
cosmology, allowing us to compare with the analytic solution.   This  
provides a test-bed for $(3+1)$ Regge calculus, whilst maintaining 
maximal freedom in the non-specifiable portion of the initial data.

The spatial length scales which appear in the geodesic expansions
above must be chosen for a particular simulation, noting that the
total volume of the initial base geometry will be $64\Delta x \Delta y
\Delta z $.   To avoid problems with the Courant condition, the
timestep is scaled together with the spatial resolution.  We choose
\begin{equation}
  \Delta t = \min\{\Delta x,\Delta y,\Delta z\} \times \Delta t_0,
\end{equation}
where $\Delta t_0 < 0.5$.
The only constraint on the choice of spatial 
length scale is the condition that we require many grid zones across
the co-ordinate horizon, which for the metric (\ref{eqn:metric}) is
located at $t$, so at $t=1$ we require $\Delta x_i<<1$
\cite{centrella86}. 

The initial-value algorithm outlined in section 4 can now be applied
to the QPL model of the Kasner cosmology.  The freely specifiable
variables are fixed using equations
(\ref{eqn:first})-(\ref{eqn:last}), with equations (\ref{eqn:shift1})
and (\ref{eqn:shift2}) being used to enforce the shift conditions.   The
remaining variables at each vertex, namely the conformal factor on
$\Sigma_0$, the four diagonal brace edges between slices, and the
three cube aligned edges in $\Sigma_1$, are calculated by solving
the available Regge equations.  The resulting system of eight
equations per vertex (a total of 1024 variables for the
QPL considered here) is solved using Newton-Raphson iteration.

Finally, we present a sample initial data solution for an axisymmetric
Kasner cosmology ($p_1=p_2=2/3,p_3=-1/3$), where we take 
$\Delta x = \Delta y = \Delta z = 0.005$ and $\Delta t = 0.001$.
The initial-value solution is found to be homogeneous to a very high
degree.  Solving the constraints for the independent variables in this
case  yields the mean values 
\begin{eqnarray}
 \psi^2         & = 1.0000000001  \nonumber \\
 \tilde  l_x^2  & = 0.0000250333  \nonumber \\
 \tilde  l_y^2  & = 0.0000250333            \\
 \tilde  l_z^2  & = 0.0000249833  \nonumber \\
         d^2    & = 0.0000177563  \nonumber 
\end{eqnarray}
where $l_x$, $l_y$ and $l_z$ are the three cube-aligned legs in
$\Sigma_{1}$,   $d$ is a diagonal brace edge between slices, and
$\psi$ is the conformal factor on $\Sigma_{0}$.
The standard deviation of both the edges and the conformal factor from
these mean values is of the order $10^{-10}$, which is the convergence
tolerance used for the Newton-Raphson iteration.

\section[]{Time evolution of the Kasner cosmology}
\label{sec:evolution}

In the previous sections we described the structure of the quantity
production lattice, and built two-surface initial data for the Kasner
cosmology.   We now consider the evolution of this data.

The initial data is evolved using the parallel Sorkin evolution scheme
described by Barrett \etal \cite{committee}.  In section
\ref{sec:kinematics} 
we constructed the simplicial lattice from the initial 3-geometry
by dragging forward individual vertices.  This procedure was used
because it results in a lattice suited to Sorkin evolution.  The key
realization in the Sorkin approach \cite{tuckey92}, based on the
original examples \cite{sorkin75}, is that when a vertex is
carried forward 
to the next slice, a new Regge equation becomes available below every
new edge created.  Thus a purely local evolution algorithm is
possible.  The evolution of a single vertex  in the quantity
production lattice creates 15 new edges, and makes available 15 Regge
equations. 

In the case of the A-type vertices, 14 brace edges
are created between $\Sigma_l$ and $\Sigma_{l+1}$, together with the
timelike edge.  This completely
encloses the region surrounding the 14 spatial edges emanating from
the vertex in $\Sigma_l$, and the timelike edge joining the vertex
to its future counterpart. The Regge equations associated with the legs in
$\Sigma_l$ emanating from the vertex A, together with the timelike
edge, may be used to solve for the lengths of the 14 new brace
edges and single new timelike edge.   A similar situation applies for
the B-type vertices, except here there are 6 new cube-aligned edges in
$\Sigma_{l+1}$,  8 new spacelike diagonal braces and a timelike leg
between the Cauchy slices.   In this case, the new Regge equations
which become available are associated with the timelike edge, the 6
cube-aligned braces below the new spatial edge, and the 8 spatial
diagonal edges in $\Sigma_l$.
Evolving a C-type vertex creates 6 new brace edges, together with 8
spatial edges and a single timelike edge, and the available Regge
equations correspond to the 6 spatial edges in $\Sigma_0$, 8
brace edges, and the timelike edge. Finally, D-type vertices 
create a vertical edge and 14 new spatial edges on the newly formed
spacelike hypersurface.  The evolution equations in this case are 
associated with the brace edges between slices and the timelike edge
at the vertex.

However, this is not the entire story.  As we saw in section
\ref{sec:dynamics}, the simplicial form of the contracted Bianchi
identities imply that four equations per vertex are dependent on the
remaining equations.   The quantity production lattice has a
total of 15 dynamical edges per vertex during evolution, but we can
only consider 11 of those as independent, with the remaining four
freely specifiable -- the simplicial equivalents of shift and lapse
freedom.

We choose to apply the lapse and shift conditions on the same
combination of edges as in the initial-value problem, however the
technique is slightly different.  Consider an A-type vertex. About
this vertex there are 6 brace edges, lying above the cube-aligned
edges in the current hypersurface.  The zero shift condition is
applied by demanding that opposing brace edges, lying along the same
``axis'', have equal lengths.  This condition is applied to each of
the three pairs of such edges.  Provided that the homogeneity of the
initial slice is maintained, this ensures that the timelike edge is
locally orthogonal to the current hypersurface.   Similar conditions
are used for the remaining vertices, applied either to brace edges
(C-type) or spatial edges on the next hypersurface (B- and D-type
vertices).   The lapse freedom is utilised to specify the
squared length of the timelike edge at each vertex.

The application of four gauge conditions per vertex leaves 11
dynamical 
edges and 15 equations at each vertex in the lattice.  Clearly 4 Regge
equations per vertex are redundant, becoming constraints which may be
tracked during evolution. The redundant equation associated with 
lapse  freedom is our ``Hamiltonian constraint'', the timelike Regge
equation.  When applying shift
conditions, we choose the redundant equation to lie below the
edges upon 
which the gauge condition is applied.    For A- and C-type vertices,
shift is applied to brace edges, so the 3 redundant equations per
vertex  correspond to the cube-aligned spatial edges in the
vertex generator (see figure \ref{fig:qpl}).  For B- and D-type
vertices, shift is applied to spatial edges in $\Sigma_{l+1}$, so the
redundant equations are associated with the brace edges lying below
the 3 cube-aligned vertex generator legs on $\Sigma_{l+1}$.

The Sorkin evolution scheme is, by
construction, a partially constrained algorithm, since the redundant 
equations associated with shift edges at A- and C-type vertices are
second order evolution-type equations, whereas at B- and D-type
vertices we discard first order 
constraint-type equations.  For simplicity we shall  refer to the
redundant shift equations at each vertex as the ``Momentum
constraints'', although they are in fact a mixture of first-order
and second-order equations.   

\begin{figure}
\centerline{\epsffile{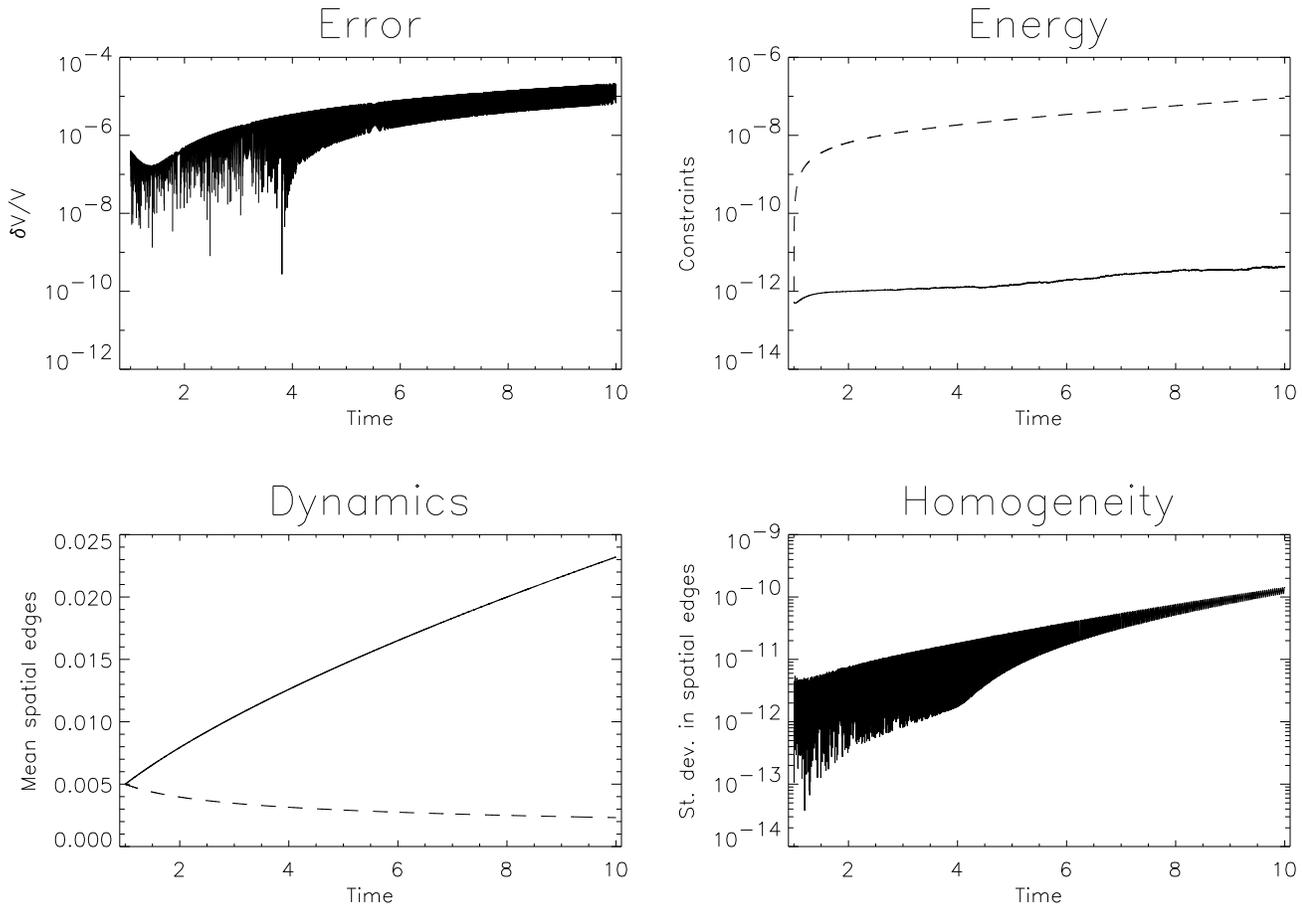}}
\caption{\protect\small{\em{The axisymmetric simplicial Kasner
solution with exponents $(2/3,2/3,-1/3)$.  The scale lengths
are taken to be $\Delta x = \Delta y = \Delta z = 0.005$, and
$\Delta t = 0.001$.  a)
The fractional error in the spatial 3-volume of the 
newly-evolved spacelike hypersurface is plotted
against the proper time measured along timelike edges in the
lattice. The error represents the fractional difference in the
simplicial 3-volume from the expected linear growth rate. 
b) Growth of the Hamiltonian (\broken) and Momentum (\full)
constraints with  time.   c) Mean value of the spatial legs $l_x$
(\full) and $l_z$ (\broken) against time.  The legs display the
expected power law behaviour, with the $l_y$ leg being identical to
$l_x$ throughout the evolution. d) Standard deviation of the spatial
leg $l_z$ from mean.  The other spatial legs $l_x$, $l_y$,  and $l_d$
behave similarly.}}} 
\label{fig:results}
\end{figure}

\begin{figure}
\centerline{\epsffile{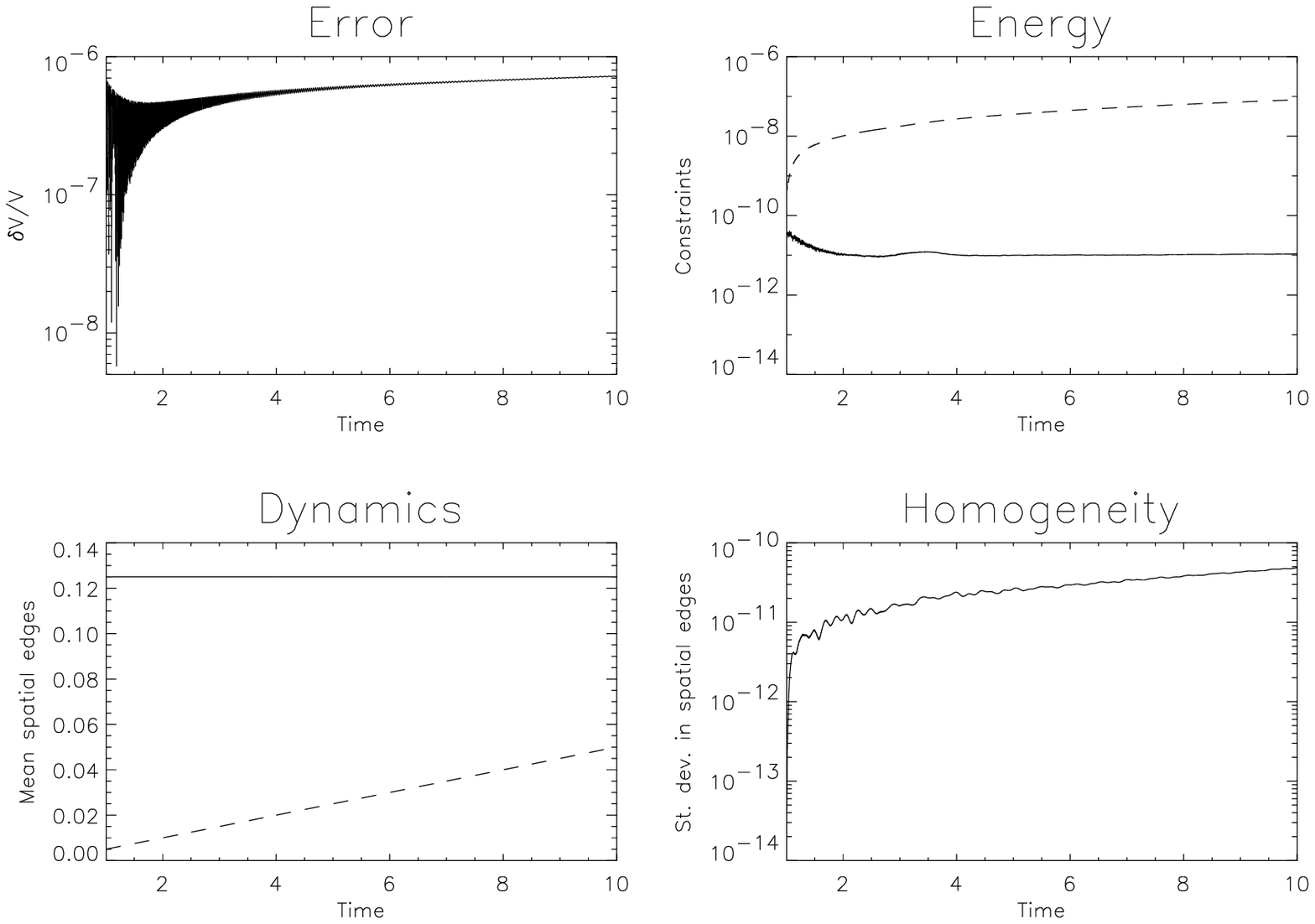}}
\caption{\protect\small{\em{The flat Kasner solution, with exponents
(0,0,1).  The lattice maintains homogeneity to
a very high degree, with the standard deviation remaining below the
limit of numerical accuracy (approximately $1\times 10^{-10}$).
The spatial and temporal scales are $\Delta x = \Delta y =
0.125$, $\Delta z = 0.005$, and $\Delta t = 0.002$.  
The same quantities are plotted as in the non-flat axisymmetric
case shown in figure \ref{fig:results}.
}}}
\label{fig:flat}
\end{figure}

The evolution of an axisymmetric Kasner cosmology (with Kasner
exponents $p_1=p_2=2/3, p_3=-1/3$) is shown in figure 
\ref{fig:results}, and figure \ref{fig:flat} shows the evolution of
an expanding flat space solution, corresponding to the Kasner exponents
$p_1=p_2=0, p_3=1$. Both runs cover a ten-fold increase in the
three-volume.
It can be seen that the fractional error in the three-volume remains
small throughout the evolution.  The fractional volume discrepancy is
defined as 
\begin{equation}   \label{eqn:volume}
   \frac{\Delta V }{V} = 1 - \frac{1}{V_0t}\sum _{^{3}V} V_i,
\end{equation}
and the summation is over all tetrahedra on a $t=constant$ 
surface.

Also apparent, particularly in the axisymmetric Kasner
solution in figure \ref{fig:results}, are high frequency oscillations
(typical wavelengths of 
a few timesteps), which decrease in magnitude as the evolution
proceeds.   It was pointed out to us by R.~Matzner that gravitational
waves in a Kasner background would evolve with such a behaviour
\cite{matzner,matzner85}.  We believe  we are seeing a grid resolution
limit effect reminiscent of such waves; however, in our case they are
not fully resolvable. The oscillations are controllable, with 
magnitudes varying as $l^2$ and wavelengths reducing as 
$l$, where $l$ is a typical edge length on the lattice.
The presence of these relatively high frequency, low  
amplitude waves does not appear to cause any instabilities, and it is
apparent from figures \ref{fig:results} and \ref{fig:flat} that the
waves gradually die out as the lattice evolves.

The evolution of the three different classes of
spatial cube-aligned edge are shown in figures \ref{fig:results}$c$ and
\ref{fig:flat}$c$, and are found to closely match the expected Kasner
evolution.   In both figures the $l_x$ and $l_y$ edges lie atop one
another.  The standard deviation of the
spatial edge $l_z$, shown in part $d$ of the figures, increases
gradually throughout the evolution.  It reaches a few parts 
in $10^{-10}$ after ten thousand timesteps for the axisymmetric Kasner
solution, and remains below $10^{-10}$ in the expanding
flat spacetime of figure \ref{fig:flat}.  

We now consider the convergence properties of the simplicial
cosmological solution.  Due to the global topology and homogeneity
of the solution, we are able to examine convergence by
reducing the typical scale length of the lattice, whilst keeping the
number of vertices fixed. This is equivalent to looking at a smaller
region of the total manifold.  For the remainder of this section we
introduce the scale parameter $\delta$, defined such that 
\begin{equation}
    \Delta x = \Delta y = \Delta z = \frac {\delta}{4}   \quad
    \hbox{and} \quad   \Delta t = 0.05 \times \delta,
\end{equation} 
which ensures that the Courant condition remains satisfied, and the
total volume of the initial base three-geometry is $\delta ^3$.  

There are several important convergence issues that must be examined
in Regge calculus, and our fully simplicial $(3+1)$ calculation
provides an excellent opportunity to do so for a specific application. 
The first, and most important issue to address is the convergence of
simplicial solutions to the corresponding solutions of Einstein's
equations.   The second issue is the consistency of the Regge
equations. That is, as the typical lattice length scale is reduced, do
the redundant Regge equations also converge?  This latter question is
related to the simplicial Bianchi identities, and the conservation of
energy momentum.  We find convergence in all solutions and equations.

\begin{figure}
\centerline{\epsffile{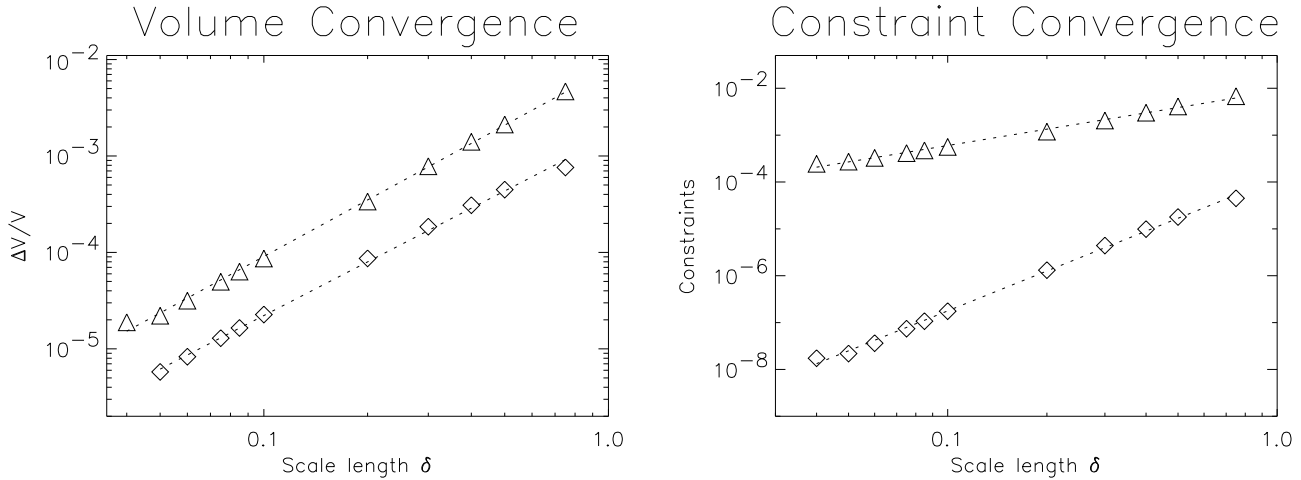}}
\caption{\protect\small{\em{Convergence estimates for the axisymmetric
Kasner model shown in figure \ref{fig:results}. a) The average
value of the fractional volume discrepancy, equation
(\ref{eqn:volume}) 
evaluated at $t=8$, is plotted against the overall scale parameter
$\delta$ ($\triangle$-points) . 
The convergence in the diagonal spatial legs ($\Diamond$-type points),
also evaluated at $t=8$, is shown.  Both the error in the volume and
the diagonal legs scales as the second power of $\delta$. This
confirms, for the axisymmetric simplicial Kasner cosmology at least,
that the 
solution of the Regge equations is a second order accurate
approximation to the continuum solution.  All other spatial legs were
found to converge as second order in the scale length.
b)  The average magnitude of the constraint equations, evaluated at
$t=8$, plotted 
against the lattice scale length $\delta$.  The momentum constraints
($\Diamond$-type points) display slightly better than second order
convergence. The Hamiltonian constraints ($\triangle$-type)
scale linearly with $\delta$.}}} 
\label{fig:converge}
\end{figure}

The error in lattice edge lengths compared to corresponding geodesic
segments in the continuum was examined, together with the fractional
difference of the simplicial three volume from linear expansion
displayed by the continuum.  The fractional volume discrepancy,
(\ref{eqn:volume}), was evaluated at $t=8$, although due to the
fluctuations apparent in figure \ref{fig:results}, the values were
averaged over several periods of the oscillation.
The results are shown in figure \ref{fig:converge}.

Figure \ref{fig:converge} also shows that the diagonal spatial edges
converge to the continuum solution second order in the typical lattice
scale length.  All other spatial edges were found to converge at
the same rate.  This indicates that the simplicial Kasner solution 
converges to the continuum solution as the second power of the
lattice spacing $\delta$.  

Recent work has cast doubt over the convergence of Regge
calculus to the continuum \cite{brewin95,miller95,reisenberger96}. 
The calculations of Brewin applied continuum geodesic
lengths directly to the 
lattice, and evaluated the Regge equations on the resulting simplicial
spacetime. Numerical convergence tests indicated that the Regge
equations failed to converge as the scale length of
the lattice was decreased.   We have shown  explicitly, for one of the
cases studied by  Brewin, that solutions of the Regge equations
converge to the corresponding solution of the continuum Einstein
equations. 

Figure \ref{fig:converge} also displays the convergence of the
redundant Regge equations as the lattice is refined.   The standard
deviation of the Momentum constraints, a combination of both
evolution and true constraint-type equations, is shown to converge as
at least the second power of $\delta$.   The Hamiltonian
constraint shows linear convergence.  The convergence of the equations
is consistent with the Regge Bianchi identities, and the existence of
diffeomorphism freedom in Regge calculus.

The convergence analysis shows that for the Kasner cosmological model,
Regge calculus is a second order discretization of Einstein's theory
of gravity.

\section[]{From BKL to Geons; future directions in Regge calculus}
\label{sec:conclusion}

We have successfully performed the first fully $(3+1)$-dimensional
calculation in Regge calculus, without the imposition of symmetry
conditions.   In the process we applied a novel 2-surface initial
value formalism to the lattice, demonstrated the gauge freedom
implied by the simplicial form of the contracted Bianchi identities,
and showed explicitly that the solution of the Regge equations
converges to its continuum counterpart. 

The simplicial Kasner solution was found to agree well with the analytic
solution, and maintains a remarkable degree of homogeneity throughout
the evolution.   Convergence analysis showed that all edge lengths
converge to their continuum values as the second power of the typical
lattice length scale, countering recent doubts over the convergence of
Regge calculus to general relativity.   The Regge constraint 
equations, arising from simplicial gauge freedoms, were also found
to converge to zero as the lattice was refined.   This demonstrates
explicitly the existence of gauge freedom in the lattice, the
simplicial counterpart of coordinate freedom in continuum general
relativity. 

Three issues direct our research in the immediate future. First 
is an analytic formulation of the convergence properties of
the Regge equations, and their solutions.  Secondly, we wish to
provide an $S^3 \times R^1$ benchmark of our code, based on
a Taub-like cosmology.  Finally, we are investigating planar
numerical perturbations \cite{centrella86} of the Kasner cosmology, in
order to analyze gravity wave propagation in our 
simplicial spacetime geometry.

In the longer term we intend to address three issues using Regge calculus.
The
first is related to the wiring of matter terms to the lattice geometry.
We cannot think of a better approach than to utilize the Cartan
analysis \cite{miller86c} described in section \ref{sec:dynamics}.
With a formulation of 
matter in Regge calculus, we can begin to study the generic properties
of collapse in inhomogeneous cosmologies.   At both early and late
stages in cosmological expansion, the effective gravity wave energy 
density dominates the contributions from matter and radiation fields.
For this reason we wish to investigate geon states
\cite{wheeler55,wheeler64} of the gravitational field
\cite{belinsky70}. We believe that these goals provide a clear
direction for the future development of Regge calculus. 

\ack
We wish to thank John A. Wheeler for his continued encouragement to
tackle this problem.  We are also indebted to Leo Brewin, Ben
Bromley, Arkady Kheyfets, Pablo Laguna, Richard Matzner and Ruth
Williams for many stimulating discussions on this and related
topics. We wish to acknowledge support for this work from an LDRD
grant from Los Alamos National Laboratory, and the Sir James McNeill
Foundation.

\section*{References}

\gdef\journal#1, #2, #3, #4 {{#1}, {\bf #2}, #3 {(#4)}. }
\gdef\FP{\it{Found.~Phys.}}
\gdef\IJTP{\it{Int. J. Theor. Phys.}}
\gdef\IJMP{\it{Int. J. Mod. Phys.}}
\gdef\GRG{\it{Gen. Rel. Grav.}}

\end{document}